\documentclass[twocolumn,secnumarabic,amssymb, nobibnotes,superscriptaddress, aps, prl]{revtex4-2}
\usepackage{graphicx}% Include figure files
\usepackage{dcolumn}% Align table columns on decimal point
\usepackage{booktabs}
\usepackage{xcolor}
\begin{document}
	
\title{Nonreciprocal Spin Waves in Nanoscale Domain Walls Detected by Scanning X-ray Microscopy in Perpendicular Magnetic Anisotropic Fe/Gd Multilayers}

\author{Ping Che}
\thanks{Present address: Unité Mixte de Physique CNRS, Thales, Université Paris-Saclay, Palaiseau 91767, France.}
\affiliation{Laboratory of Nanoscale Magnetic Materials and Magnonics, Institute of Materials (IMX), \'Ecole Polytechnique F\'ed\'erale de Lausanne (EPFL), 1015 Lausanne, Switzerland}
\author{Axel Deenen}
\affiliation{Laboratory of Nanoscale Magnetic Materials and Magnonics, Institute of Materials (IMX), \'Ecole Polytechnique F\'ed\'erale de Lausanne (EPFL), 1015 Lausanne, Switzerland}
\author{Andrea Mucchietto}
\affiliation{Laboratory of Nanoscale Magnetic Materials and Magnonics, Institute of Materials (IMX), \'Ecole Polytechnique F\'ed\'erale de Lausanne (EPFL), 1015 Lausanne, Switzerland}
\author{Joachim Gr\"afe}
\affiliation{Max Planck Institute for Intelligent Systems, Heisenbergstraße 3, 70569 Stuttgart, Germany}
\author{Michael Heigl}
\affiliation{Institute of Physics, University of Augsburg, Universit\"atsstrasse 1, D-86159 Augsburg, Germany}
\author{Korbinian Baumgaertl}
\affiliation{Laboratory of Nanoscale Magnetic Materials and Magnonics, Institute of Materials (IMX), \'Ecole Polytechnique F\'ed\'erale de Lausanne (EPFL), 1015 Lausanne, Switzerland}
\author{Markus Weigand}
\affiliation{Helmholtz-Zentrum Berlin für Materialien und Energie, Albert-Einstein-Straße 15, 12489 Berlin, Germany}
\author{Michael Bechtel}
\affiliation{Helmholtz-Zentrum Berlin für Materialien und Energie, Albert-Einstein-Straße 15, 12489 Berlin, Germany}
\author{Sabri Koraltan}
\affiliation{Physics of Functional Materials, Faculty of Physics, University of Vienna, Kolingasse 14-16, A-1090, Vienna, Austria}
\affiliation{Vienna Doctoral School in Physics, University of Vienna, Kolingasse 14-16, A-1090, Vienna, Austria}
\author{Gisela Sch\"utz}
\affiliation{Max Planck Institute for Intelligent Systems, Heisenbergstraße 3, 70569 Stuttgart, Germany}
\author{Dieter Suess}
\affiliation{Physics of Functional Materials, Faculty of Physics, University of Vienna, Kolingasse 14-16, A-1090, Vienna, Austria}
\affiliation{Research Platform MMM Mathematics-Magnetism-Materials, University of Vienna, Vienna 1090, Austria}
\author{Manfred Albrecht}
\affiliation{Institute of Physics, University of Augsburg, Universit\"atsstrasse 1, D-86159 Augsburg, Germany}
\author{Dirk Grundler}
\email{dirk.grundler@epfl.ch}
\affiliation{Laboratory of Nanoscale Magnetic Materials and Magnonics, Institute of Materials (IMX), \'Ecole Polytechnique F\'ed\'erale de Lausanne (EPFL), 1015 Lausanne, Switzerland}
\affiliation{Institute of Electrical and Micro Engineering (IEL), \'Ecole Polytechnique F\'ed\'erale de Lausanne (EPFL), 1015 Lausanne, Switzerland}
\date{\today}

\begin{abstract}
Spin wave nonreciprocity in domain walls (DWs) allows for unidirectional signal processing in reconfigurable magnonic circuits. Using scanning transmission x-ray microscopy (STXM), we examined coherently-excited magnons propagating in Bloch-like DWs in amorphous Fe/Gd multilayers with perpendicular magnetic anisotropy (PMA). Near 1 GHz we detected magnons with short wavelengths down to $\lambda = 281$~nm in DWs whose minimum width amounted to $\delta_{\rm DW} = 52$~nm. Consistent with micromagnetic simulations, the STXM data reveal their nonreciprocal magnon band structures. We identified Bloch points which disrupted the phase evolution of magnons and induced different $\lambda$ adjacent to the topological defects. Our observations provide direct evidence of nonreciprocal spin waves within Bloch-like DWs, serving as programmable waveguides in magnonic devices with directed information flow. 
\end{abstract}

\maketitle

\newpage
%Introdcution
Spin waves, the quantized collective excitation of magnetic moments, can exhibit nonreciprocity in amplitude and/or frequency $f$ when propagating with wave vectors $\mathbf{k}$ in opposite directions. It is of great importance for magnonic devices such as the isolators, circulators and the spin diodes \cite{Chumak_review}. Damon-Eshbach modes possess symmetric dispersion relations $f(k)$ but display asymmetric amplitudes due to the gyrotropy of spin precession \cite{Damon-Eshbach, Ciubotaru2016, Devolder2023}. Spin wave dispersion relations which are asymmetric in $k$ lead to frequency nonreciprocity. This nonreciprocity can be engineered by Dzyaloshinskii–Moriya interaction (DMI) \cite{Moon2013,Di2015,Kuess2023,Cortes2013,Seki2016,Sato2016,Che2021}, the chirality of magneto-dipolar interactions between magnetic thin films and nanomagnets \cite{Liu2018,Chen2019,Wang2021,Temdie2023}, the curvature of a three-dimensional nanostructure \cite{Hertel2013,Otalora2016,Giordano2023} or magnetic coupling between two magnetic layers \cite{Gallardo2019,Ishibashi2020,Kuess2021}. Nonreciprocity of spin waves in domain walls (DWs) is particularly interesting for the possibility of curved short-waved spin waves transmission. Wagner et al. \cite{Wagner2016} suggested nanomagnonic circuits based on reconfigurable DWs in thin films, but relevant DW widths and spin-wave nonreciprocity were not experimentally addressed. In Ref. \cite{Garcia-Sanchez2015} Néel-type DWs in a metallic ferromagnet adjacent to a heavy metal were considered and non-reciprocal properties were predicted. However, such materials systems with interfacial DMI are known to exhibit an increased spin-wave damping. Alternatively, Bloch-type DWs which channel Winter-mode spin waves \cite{Winter1961,Sluka2019} are expected to exhibit asymmetric dispersion relations as well. They are a result of dynamic dipolar interactions and have been modelled by micromagnetic simulations. \cite{Henry2016,Henry2019,Chen2020}. Still, an experimental verification has so far remained elusive because DWs stabilized in a material with low damping, coherent spin wave excitation at GHz frequencies and a phase-coherent imaging technique with a high-spatial resolution are all required in the same setup.\\
\begin{figure*}
    \centering
    \includegraphics[width=178mm]{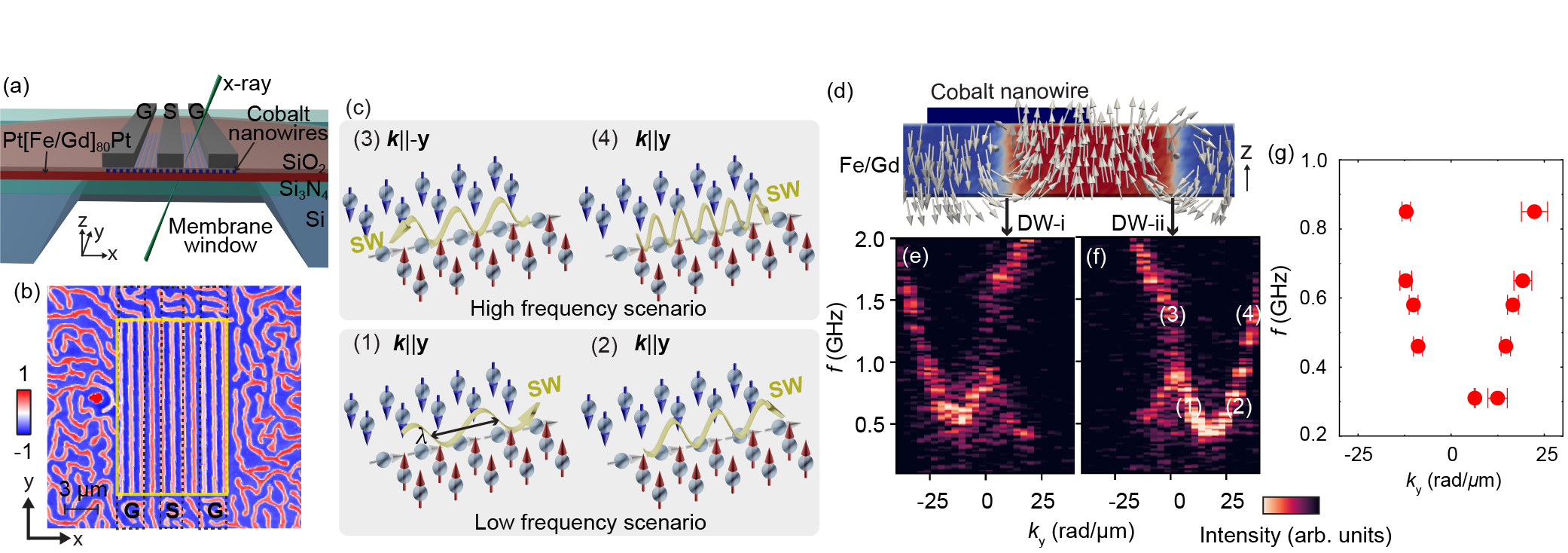}
    \caption{(a) Schematic diagram of the device and measurement configuration of STXM through the membrane window. CPW consists of the ground (G) and signal (S) lines for spin wave excitation. (b) Static STXM images of domains and DWs modified by 1D Co nanowire arrays of periodicity $p_\mathrm{nw}$ = 450 nm at $\mu_0 H_{\perp}$ = 0 mT. The color bar represents normalized x-ray transmission intensity. Yellow frame indicates the region of the Co arrays and the region outside shows the disordered stripe domains in the bare Fe/Gd multilayers. Dashed black frames indicate the regions of the CPW. (c) Two scenarios of spin waves propagation with wavelength $\lambda$ in the dispersion relations depicted in (f). Low frequency scenario consists of (1) and (2). High frequency scenario consists of (3) and (4). Yellow arrows describe the phase velocity directions. (d) Simulated DW structure in the Fe/Gd multilayers. The red and blue colors reflect the spin orientations along the $\textbf{z}$-axis. The DW-i is underneath the Co nanowire. DW-ii is in the gap between nanowires. (e) and (f) Asymmetric spin wave dispersion relations corresponding to the Bloch part of the DW-i and DW-ii extracted from the center layer along $\textbf{z}$-axis in micromagnetic simulation. (g) Dispersion relations extracted from the STXM imaged spin waves within the DWs located at the gap of nanowires like a DW-ii.}
    \label{fig1}
\end{figure*}
\indent
In this letter, we employ scanning transmission x-ray microscopy (STXM) for imaging statically the magnetic configuration in Fe/Gd multilayers and dynamically the spin waves excited by an integrated coplanar waveguide (CPW). STXM provides a spatial resolution of 20 nm and a temporal resolution of 50 ps \cite{Wintz2016,Baumgartner2017,Finizio2018,Finizio2019,Albisetti2020,Trager2021_PRB}. The amorphous Fe/Gd multilayers exhibit a perpendicular magnetic anisotropy (PMA) and low damping $\alpha$ of about 10$^{-3}$ \cite{Montoya2017_damping}. Different spin textures have been reported in such multilayers such as stripe domains, dipole skyrmions, and antiskyrmions if combined with Ir \cite{Lee2016,Montoya2017_phase,Montoya2018,Desautels2019,Heigl2021}. Here, we focus on DWs which are of the Bloch-like type. We have prepared Cobalt (Co) nanowire arrays on top of the Fe/Gd multilayers grown on Si$_3$N$_4$ membrane and stabilized a preferred domain alignment. Utilizing STXM, we identify the nonreciprocal characteristics of the spin waves within the DWs excited by the torque provided by radio-frequency (rf) currents in CPW. Depending on the frequency we observe two different signatures of nonreciprocity. At low frequencies, spin waves exhibit the same sign of the phase velocity $v_\mathrm{p}=2\pi f/k$ (and the wave vector) but opposite sign of group velocities $v_\mathrm{g}$. At higher frequencies, spin waves exhibit wave vectors with different signs and magnitudes as further substantiated by micromagnetic simulations. A topological defect by a Bloch point does not modify the dispersion relation but disrupt the phase evolution of the propagating spin waves. \\
\indent
%Technique description and Sample description
The cross section of the device is sketch in Fig. \ref{fig1}(a). Co nanowire arrays and coplanar waveguide (CPW) consisting of signal line (S) and ground (G) lines were fabricated on amorphous [Gd(0.5 nm)/Fe(0.35 nm)]x80 multilayers and the sample preparation is described in Supplementary Materials (Supp.) section I \cite{Supp}. Two samples A and B containing multiple devices will be discussed. We performed static and dynamic x-ray circular dichroism (XMCD) measurements using the STXM at MAXYMUS station, Bessy II, Berlin \cite{Grafe2019} at the Gd M5 edge 1187.6 eV. The devices were tilted to 30 degree with respect to the incident x-ray as shown in Fig. \ref{fig1}(a) in order to collect both the in-plane and out-of-plane dynamic magnetization. A converting factor 2/$\sqrt{3}$ was applied to obtain the actual $\textbf{x}$-axis pixels. Before performing the STXM measurements, the Co nanowires were magnetized by an in-plane magnetic field along their long axes and then the samples were demagnetized with an oscillating out-of-plane field. An external field $\mu_0\textbf{H}_{\perp}$ was applied perpendicular to the sample surface to stabilize the investigated magnetic configuration. \\
\begin{figure*}
    \centering
    \includegraphics[width=178mm]{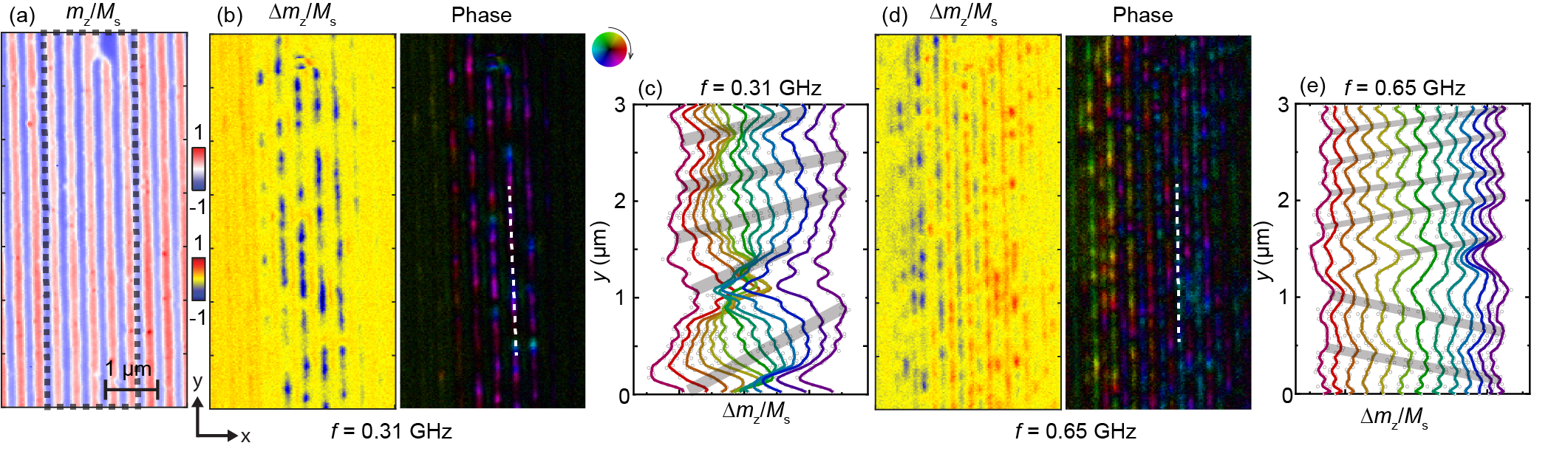}
    \caption{(a) Static STXM images of the densest domains formed in Fe/Gd multilayers underneath and near the signal line. Here 1D Co nanowire array of periodicity $p_\mathrm{nw}$ = 350 nm at $\mu_0 H_{\perp}$ = 0 mT is integrated. Grey frames indicated the region covered by the signal line. (b) and (d) Snapshot of the spin dynamics and their phase (integrated with the amplitude) at $f$ = 0.31 GHz and $f$ = 0.65 GHz normalized to the static image. (c) and (e) The time evolution of the dynamics components of the transmission signal taken from the region marked by the dashed white lines in all the phase images. Grey shadows are the eye-guide by marking the maxima moving with time.}
    \label{fig2}
\end{figure*}
\indent
%Static imaging
The static STXM image of sample-A containing a nanowire array with stripes of periodicity $p_\mathrm{nw}$ = 450 nm and a width of $w_\mathrm{nw}$ = 225 nm is depicted in Fig. \ref{fig1}(b). The external field $\mu_0 H_{\perp} = 0$~mT was applied along $\textbf{z}$-direction. The normalized transmitted x-ray signal is displayed to imitate $m_\mathrm{z}$/$M_\mathrm{S}$. The integrated CPW (marked by grey frames) modified locally the magnetic contrast because it absorbed part of the x-ray and reduced the transmission. The well-aligned stripe domains and DWs were near the Co nanowire arrays. The periodicity of the domain lattice $p_\mathrm{d}$ is $2p_\mathrm{nw}$. The bare Fe/Gd outside the nanowire region contained the random domains. In sample-B with slightly different materials parameters, denser domain lattices were produced which fulfilled $p_\mathrm{d} = p_\mathrm{nw} = 350$~nm [Fig. \ref{fig2}(a) and Supp. Fig. S1(b)]. STXM images of devices with different nanowire sizes on sample-B in Supp. Fig. S1 \cite{Supp} showed an evolution of magnetic states from partially aligned domains to domains fully aligned with nanowires. \\ 
\indent
%The domain state and the dispersion relation by simulation
Micromagnetic simulations were performed using the Mumax3 code \cite{Mumax3,Mumax3_thermal,temperature_mesh,abs_boundary} to understand the magnetic domain structures and their spin dynamics. The static and dynamic simulation parameters are described in Supp. section III. The PMA and the dipole interaction cooperated to construct the DWs in Fe/Gd multilayers and the spin textures at top and bottom surfaces experienced Néel-type rotation while the central part was Bloch-type [Fig. \ref{fig1}(d)]. Later, we will distinguish two kinds of DWs considering the surrounding domains extending into $y$-direction: spin-down on the left and spin-up on the right (labelled as DW-i), and spin-up on the left and spin-down on the right (labelled as DW-ii). Figure \ref{fig1}(e) and (f) describe the simulated asymmetric spin wave dispersion relations inside the Bloch-type regions of these two DWs. The variation of the dispersion relations over the thickness are displayed in the Supp. Fig. S2 \cite{Supp} and they are consistent in terms of nonreciprocity. There are two typical scenarios of nonreciprocity in each dispersion relation. They are illustrated in Fig. \ref{fig1}(c) considering the DW-ii and the dispersion relation $f(k_y)$ shown in Fig. \ref{fig1}(f). At low frequency, spin wave excitation occurs exclusively when $k_{y}>0$. The phase velocity $v_\mathrm{p}$ is always positive. Two modes are excited at the same time, (1) one at small $k$ with $v_\mathrm{p}<0$ and (2) one at large $k$ with $v_\mathrm{p}>0$. At high frequency, modes exist with $k_{y}>0$ and $k_{y}\leq 0$. Here, both modes have positive $v_\mathrm{g}$. Our simulation results are consistent with the modelling and simulations performed on Bloch-like DWs in Refs. \cite{Henry2016,Henry2019}.\\
\indent
%Dynamic STXM
Figure \ref{fig2} shows STXM data obtained on domains in device2 of sample-B underneath and near the signal line. In Fig. \ref{fig2}(a) the regular stripe-domain patterns is shown which is formed below the Co nanowire array with $p_{\rm nw}=350$~nm. The domains are about 120 nm wide and appear in red and blue color indicating regions with out-of-plane magnetization vectors pointing up and down, respectively. The DWs are white indicating in-plane magnetization vectors with width of $(60\,\pm\,13)$~nm. We applied continuous-wave rf currents at multiple frequencies and excited the depicted domain pattern. The dynamic magnetization $\Delta m_z(y)$ (spin-precessional amplitude) was stroboscopically detected with high temporal resolution using a pulsed x-ray beam. In Fig. \ref{fig2}(b) a snapshot of the normalized spin-precessional amplitudes at $f = 0.31$~GHz and the local phases (phase map) are plotted. Attributable to the DW alignment, the signal in the DW marked by the dashed white line is further examined in Fig. \ref{fig2}(c) where we show the time evolution of the dynamic magnetization $\Delta m_z(y)/M_{\rm s}$ as a waterfall plot. The grey bars highlight how local maxima in $\Delta m_z(y)/M_{\rm s}$ shift in space with time. They hence define the local wavelength. It is notable that below and above $y = 1.5~\mu$m, these maxima show both a different separation in $y$ and different temporal shift. Below [above] $y = 1.5~\mu$m we extract a wavelength $\lambda_1 = (988 \pm 26)$~nm [$\lambda_2 = (504 \pm 105)$ nm] with $v_\mathrm{p}> 0$, propagating along the $+y$-direction (Supplemental Movie1). The observation agrees with the low frequency scenario illustrated in Fig. \ref{fig1}(c).\\ 
\indent 
In Fig. \ref{fig2}(d) and (e) we analyze the spin dynamics of the same DW at an increased excitation frequency of $f = 0.65$~GHz. Below [above] $y = 1.5~\mu$m we now extract a wavelength $\lambda_1 = (515 \pm 66)$~nm with $v_\mathrm{p}<0$ [$\lambda_2 = (327 \pm 41)$~nm with $v_\mathrm{p}>0 $]. The two spin waves propagate into opposite directions along the $y$-axis (Supplemental Movie2), consistent with the high frequency scenario in Fig. \ref{fig1}(c). Further STXM images at $f = 0.46$~GHz and 0.58~GHz are shown in Supp. Fig. S3 \cite{Supp}. In Fig. \ref{fig1}(g) we summarize the observed wavelengths and phase velocities. Short-waved spin waves down to $\lambda = (281 \pm 44)$~nm were channelled in the DWs at $f$ = 0.85 GHz. This value is more than a factor of $10^{6}$ times shorter than the wavelength $\lambda_{\rm em}$ of the corresponding electromagnetic wave in free space and, to our knowledge, a record on-chip miniaturization of $\lambda_{\rm em}$. It substantiates the prospects of nanomagnonics based on domain walls and noncollinear spin structures \cite{Wagner2016,Duerr2012}. The experimental data provide the dispersion relation $f(k_y)$ inside the investigated DW. We find a clear asymmetry between the spin wave branches at positive and negative wave vectors $k_y$. The experimentally resolved nonreciprocity for spin waves in the DW agrees qualitatively well with the characteristics extracted from micromagnetic simulations [Fig. \ref{fig1}(f)]. The remaining discrepancy concerning measured and simulated eigen-frequencies might be caused by the modification of the magnetic properties of the Fe/Gd multilayers during the nanofabrication. The lithography for lift-off processing of Co nanowires and CPWs involved processes at elevated temperatures which facilitated partial interdiffusion of Fe and Gd.\\
\begin{figure}
    \includegraphics[width=86mm]{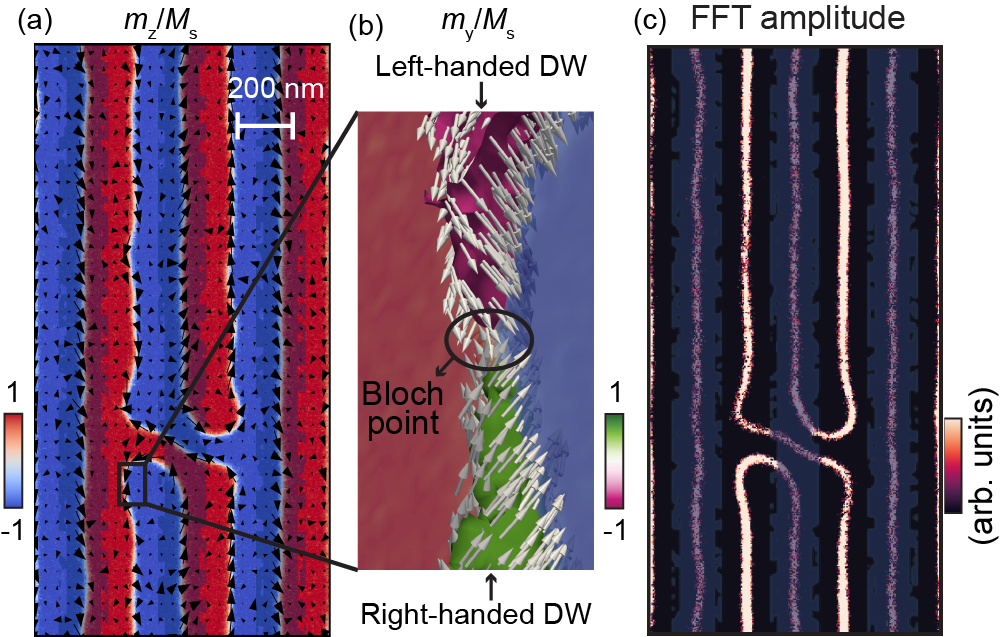}
    \caption{(a) Static magnetization $m_\mathrm{z}$/$M_\mathrm{S}$ of the domain configuration extracted from micromagnetic simulations in the layer number 12 along z-direction of Fe/Gd film. Black arrows reflect the in-plane magnetization directions. (b) Zoom of (a) with isosurfaces of $m_{\rm y}/M_{\rm S}=\pm~0.75$, showcasing domain walls with opposite handedness separated by a Bloch point. The color bar indicates $m_\mathrm{z}$/$M_\mathrm{S}$. (c) Simulated spin dynamics amplitude at $f = 0.42$~GHz. The legend indicates the logarithmic scale of the normalized Fourier component of the magnetization in resonance. Blue shadows in (a) and (c) indicate the Co nanowires with defects.}
    \label{fig3}
\end{figure}
\indent
Near $y=1.5\,\mu$m in Fig. \ref{fig2}(c) and (e), a spin pinning effect affects the spin dynamics. To gain microscopic insight, we carried out dynamic simulations for a DW containing a Bloch point [grey rectangle in Fig. \ref{fig3}(a)]. Near this point, the in-plane magnetization vectors rotate by 180 degrees [Fig. \ref{fig3}(b)]. As a consequence, the handedness of the DW switches locally. This variation cannot explain the two different spin waves above and below $y=1.5~\mu$m as the handedness does not modify the dynamic dipolar interaction; the dispersion relation in the Bloch-type DW stays the same \cite{Henry2019}. We assume that the non-collinear spin structure at the Bloch point operates as a point-like scatterer upon microwave excitation allowing us to emit spin waves propagating in opposite directions inside the DW at high frequency.\\
\begin{figure}
    \includegraphics[width=78mm]{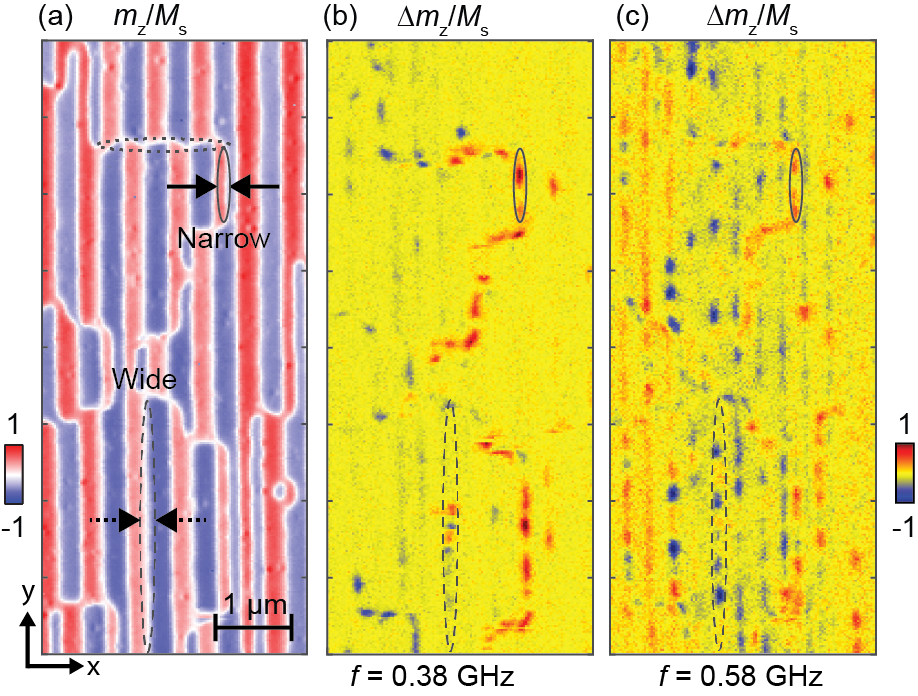}
    \caption{(a) Static STXM image of domains formed by Co nanowire arrays with $p_\mathrm{nw}$ = 300 nm. Solid arrows and circle mark the example of a narrow DW. Dashed arrows and circle mark wide DW. Dotted circle marks the example of DW segments along x-direction. (b) and (c) Snapshots of the DWs dynamics at $f$ = 0.38 GHz and $f$ = 0.58 GHz, respectively.}
    \label{fig4}
\end{figure}
\indent
In Fig. \ref{fig2} and Supp. Fig. S3 we observed that at the small frequencies $f = 0.31$~GHz and 0.46 GHz not all DWs were excited. Instead, spin waves existed particularly in the DWs-ii. The same behavior was reproduced by our micromagnetic simulations [Fig. \ref{fig3}(c)]. This is different for the higher excitation frequencies for which both DWs-ii and DWs-i hosted spin waves. At $f=0.58$~GHz, two spin waves modes of $\lambda = (739 \pm 85)$~nm and $\lambda= (803 \pm 141)$~nm with $v_\mathrm{p}<0$ were detected by STXM, consistent with the low frequency scenario in the dispersion relation shown in Fig. \ref{fig1}(e). This observation indicated that the dispersion relation in DWs-i is blue shifted compared to DWs-ii. We noted that this discrepancy existed underneath the Co nanowire arrays and vanished for DWs in a bare Fe/Gd multilayer as shown in the dynamic STXM images of Supp. Fig. S4. In bare Fe/Gd, spin waves were channeled in both kinds of DWs which corroborated that the Co nanowires introduced the frequency-dependent selection rule. In the framework of the equation of motion (Landau-Lifshitz equation) \cite{Gurevich}, one might assume that the DWs-i exhibited an enlarged internal field $H_{\rm int}$. A detailed evaluation of the quasi-static internal fields extracted from the simulations did not display a clear discrepancy in $H_{\rm int}$ of the two different DWs. We attribute the observed discrepancy in DW characteristics to an asymmetry in dynamic coupling between the magnetization vector of Co nanowires and the two type of DWs directly underneath.\\ 
\indent
In Fig. \ref{fig4}(a) we show an STXM image taken on an Fe/Gd multilayer underneath a Co nanowire array with a small periodicity $p_\mathrm{nw}= 300$~nm (sample-B device1). We observe a particular domain configuration with different types of DWs, which we call super-domain structure in the following. The majority of domains have a width of about 220 nm. Between such domains we find a DW along $y$-direction [dotted oval in Fig. 4(a)] which we label wide-domain DW (w-DW). Its width amounts to ($83\,\pm\,21$) nm. The combined width of a domain and a DW is commensurable with $p_{\rm nw}$ in such a pattern. A few domains are locally narrower than 220 nm as additional DWs exist, highlighted by the gray oval in Fig. \ref{fig4}(a) and labelled narrow-domain DW (n-DW). We find that the n-DWs are located in gaps between Co nanowires. Their width amounts to ($52\,\pm\,10$) nm. Furthermore, there are DW segments which extend along the $x$-direction. Their width amounts to ($104\,\pm\,17$) nm. They connect to both w-DWs and n-DWs and thereby produce a super-domain structure consisting of patches of regular 300-nm-wide stripe domains. Figure \ref{fig4}(b) displays a snapshot of the dynamic STXM images taken for an excitation frequency $f = 0.38$~GHz. Prominent spin-precessional motion (dark red and dark blue) is found only in a few narrow channels which follow the curved boundaries defining the super-domain structure. Inside the patches of regular 300-nm-wide domains the excitation is weak. A similar distribution of spin-precessional amplitudes is found at $f = 0.46$~GHz (not shown). The detailed analysis reveals that channels along the $y$-axis host propagating spin waves, while along the $x$-axis standing spin waves exist.\\ 
%A similar distribution of spin-precessional amplitudes is found at $f = 0.46$~GHz (not shown).
\indent 
At $f=0.58$~GHz, the distribution of spin-precessional amplitudes is found to be completely different [Fig. \ref{fig4}(c)]. Spin waves are now channeled inbetween many more wide and narrow domains. In the dashed oval, we find two propagating spin wave modes with $v_\mathrm{p}>0$ which exhibit wavelengths $\lambda= (672 \pm 80)$~nm and $\lambda = (395 \pm 72)$~nm. The observation is consistent with the low-frequency scenario of Fig. \ref{fig1}(c). The experimental data obtained on device1 further corroborate the nonreciprocal spin wave dispersion relations inside DWs in Fe/Gd multilayers. We note that spin waves propagating along the $y$-direction are counter-intuitive as their wave vectors are collinear with the direction of the rf current in the CPW. We assume that a defect or again a Bloch point inside the DW allowed us to emit spin waves into $y$-direction.\\
\indent
In summary, we presented the experimental exploration of nonreciprocal spin wave dispersion relations in Bloch-like DWs in Fe/Gd multilayers by dynamic STXM measurements. The integration of Co nanowire arrays stabilized domain patterns and DWs. Depending on their periodicity we found periodic stripe domains or a distinct super-domain structure. We demonstrated that the presence of topological singularities known as Bloch points did not alter the nonreciprocity of spin waves. Still, they affected their phases. Depending on the excitation frequency, we realized uni-directional flow of spin waves exhibiting two characteristic wavelengths determined by the nonreciprocal spin-wave dispersion relation. Our experiments indicate that integrated nanomagnets can be used to imprint super-domain structures and complex DW configurations which channel spin waves through Fe/Gd multilayers. The minimum channel width amounts to about 50 nm. By intentionally introducing defects at controlled locations, one can further control the phase of propagating spin waves. Our findings promise the design of magnonic logic circuits at GHz frequencies which make use of nonreciprocal magnon band structures in ultra-narrow spin-wave channels formed by DWs in PMA thin films.\\
\indent
The authors acknowledge the financial support from the Swiss National Science Foundation (SNSF) via grants 197360 and CRSII5 171003, as well as the Deutsche Forschungsgemeinschaft (DFG) via TRR80: From Electronic Correlations to Functionality. Measurements were conducted at the Maxymus endstation at BESSY2, Helmholtz-Zentrum Berlin (HZB), Berlin, Germany. The authors thank HZB for the allocation of synchrotron radiation beamtime. The simulations and post-processing have been performed using the facilities of the Scientific IT and Application Support Center of EPFL.

\end{document}